\documentclass[referee]{aastex} 

\usepackage{graphicx}
\usepackage{natbib}
\usepackage{ulem}
\usepackage{txfonts}

\slugcomment{Submitted to the Astrophysical Journal}

\shorttitle{Statistical analysis of the very quiet Sun magnetism}
\shortauthors{Mart\' inez Gonz\'alez et al.}

\begin{document}

\title{Statistical analysis of the very quiet Sun magnetism}

\author{M.\ J.\ Mart\' inez Gonz\'alez, R. Manso Sainz, A. Asensio Ramos}
\affil{Instituto de Astrof\'{\i}sica de Canarias, C/V\'{\i}a L\'actea s/n, 38200
La Laguna, Tenerife, Spain\\
Departamento de Astrof\'{\i}sica, Univ. de La Laguna, 38205, La Laguna, Tenerife, Spain}
\email{marian@iac.es}
 
\and
 
\author{A. L\'opez Ariste}
\affil{THEMIS-CNRS UPS 853, V\' ia L\'actea S/N, 38200, La Laguna, Tenerife,
Spain}

\begin{abstract}
The behavior of the observed polarization amplitudes with spatial resolution is a strong
constraint on the nature and organization of solar magnetic fields below the resolution limit.
We study the polarization of the very quiet Sun at different spatial resolutions using
ground- and space-based observations.
It is shown that 80\% of the observed
polarization signals do not change with spatial resolution, suggesting
that, observationally, the very quiet Sun magnetism remains the same despite the high spatial resolution
of space-based observations.
Our analysis also reveals a cascade of spatial scales
for the magnetic field within the resolution element.
It is manifest that the Zeeman effect is sensitive to the microturbulent field usually
associated to Hanle diagnostics. This demonstrates that Zeeman and Hanle studies show
complementary perspectives of the same magnetism. 
%
\end{abstract}


\section{Introduction}
The quiet Sun comprises all areas outside active regions in the solar
surface, which, it turn, is formed by the network and the very quiet areas inside it,
the so-called internetwork. Internetwork magnetism is characterized by an
intermittent pattern of positive and negative polarities, revealing a complexity
larger than the one present in active regions or even in the network. The presence
of a scale cascade for bipolar regions was already suggested by \cite{stenflo92}. This
cascade apparently goes down to the sub-arcsecond or ``turbulent'' field that permeates
99\% of the photosphere that is not occupied by the kG flux tubes of the network. 
It is now clear that internetwork magnetic fields are formed by a bunch of mixed-polarity
magnetic fields coexisting in a single resolution element \citep[e.g.][]{jorge_egidio_valentin_96, lites_02, khomenko_03, hector_04, ita_06, david_07, marian_08}. However,
all these works had to apply very simplistic models to answer a fundamental question: which 
are the strengths of internetwork magnetic fields?.
This question is of paramount importance to estimate the role of the internetwork
on the photospheric magnetism, specially because of the large area that it occupies.

Models used to infer physical information from the observed Zeeman signals are
usually based on one (or two) magnetic atmosphere(s) embedded in a non-magnetized plasma.
This strategy has allowed to put in evidence that internetwork magnetic fields have a
preference for hG field strengths and suggests that the topology of the magnetic
field is quasi-isotropic \citep{marian_08, marian_andres_08, andres_09, bommier_09}
\footnote{Note that the
overabundance of horizontal fields found in \cite{david_07} or \cite{lites_08}
is roughly compatible with an isotropic distribution of field vectors, in which the
cosine of the inclination is uniformly distributed.}. 

Diagnostic techniques based on the Hanle effect
model the internetwork magnetic field assuming that it is microturbulent
(isotropic below the mean free path of line-core photons).
Within this scenario, magnetic field stregths $\sim 10$-100~G \citep{stenflo82},
and $\sim 20$-30~G \citep{faurobert_01}, have been inferred.
A more recent estimate \citep{javier_04}, based on three-dimensional radiative
transfer calculations and state-of-the-art atmospheric models finds 
mean field strengths as high as $\langle B\rangle\sim 130$~G. Evidence that Hanle depolarization by a randomly oriented field is
indeed at work in the quiet solar photosphere was provided by
\cite{rafa_04}, though without giving strong constrains on the actual value of $\langle B\rangle$.

Simple models favor a straightforward interpretation of the data and provide an
estimate of a (non-linear) average of the distribution of field strengths in the very quiet Sun.
Unfortunately, since the field is known to be organized below the resolution element,
this structuring escapes to our detection. Obviously, any simplified modeling fails
at extracting all the information from such a complex region of the photosphere. One
should rely, instead, on a statistical description of the internetwork magnetism 
\citep[e.g., ][]{jorge_egidio_valentin_96, thorsten_markus_07, thorsten07}.

In this letter, we aim at investigating the nature of internetwork magnetism
by inferring the characteristic sizes of magnetic elements. We analyze if they are macroscopic structures,
microturbulent magnetic fields, or if different spatial scales coexist in the
same resolution element. The variation with the spatial resolution of the
polarization amplitudes we observe with the Zeeman effect gives strong constrains 
to the nature and organization of these magnetic fields.

\section{Observations}
We analyze high sensitivity quiet Sun spectro-polarimetric data of the Fe\,{\sc i} 
lines at 630 nm using different instruments. The spectropolarimeter \citep[SP;][]{litesetal_01}
aboard Hinode spacecraft \citep{kosugi_hinode07} provides us with high spatial resolution data (0.32$''$). The observations consist of a time series of a fixed slit position at disk
center acquired on February 27 2007. The noise level in the polarization profiles is $2.9\times 10^{-4}$ in
terms of the continuum intensity, $I_\mathrm{c}$. More details can be found in
\cite{lites_08}.

Two other data sets having a spatial resolution of the order of 1.3$''$ were obtained
on August 17, 2003 using the POLIS instrument \citep{beck_05} attached to the VTT and 
on July 5 2008 using the ZIMPOL \citep{gandorfer04}
instrument mounted at TH\'EMIS, both telescopes at El Teide observatory. The POLIS data consist of a map of 33.25$'' \times$ 42$''$, 
along the slit and scan directions,
respectively. The integration time at each slit position was $\sim$27~s, which
allowed a noise level in the polarization profiles of \hbox{$7\times 10^{-5}$
I$_\mathrm{c}$}. Details about data reduction and the observations are
given by \cite{marian_08}. The ZIMPOL data consist of a time series of a small 
scan of a quiet solar region at
disk center. The scanned region was \hbox{$73.7'' \times 5''$} along the slit
and the scan direction, respectively. The spatial resolution was $\sim 1.3''$ and the noise 
level in the polarization profiles is of about
\hbox{$5\times 10^{-4}$ I$_\mathrm{c}$} after the denoising using the procedure
described by \cite{marian_08}.

\begin{figure}[!t]
\centering
\includegraphics[width=\columnwidth]{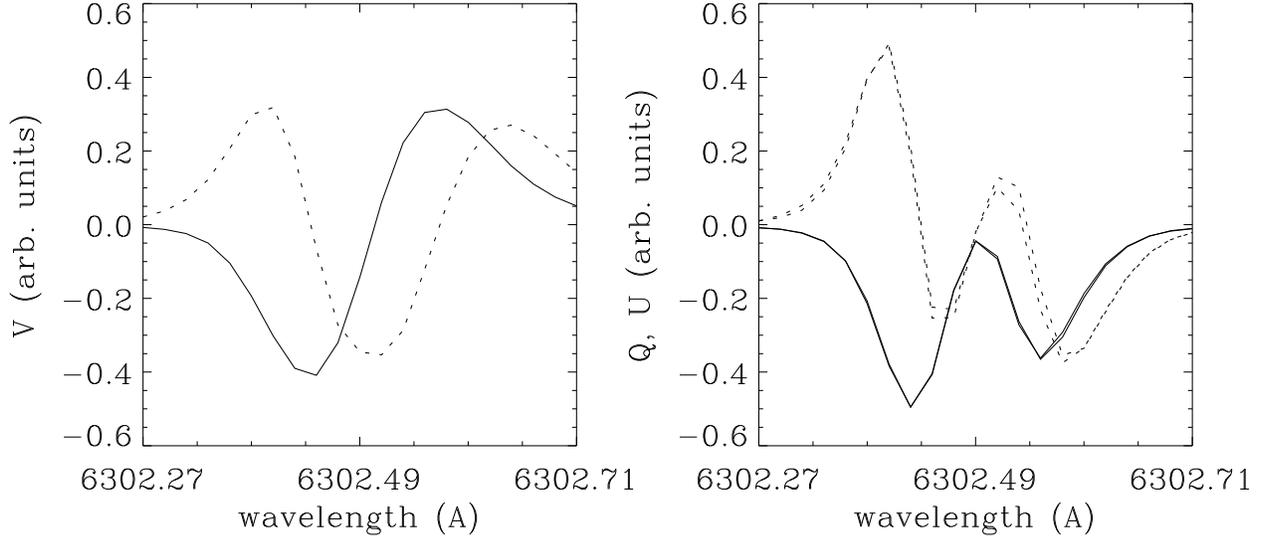}
\caption{First (solid line) and second (dashed lines) principal component of the HINODE Stokes
$V$ (left panel) and $Q$ and $U$ (right panel) obtained from the observational data. 
Note the similarity between the Stokes $Q$ and $U$ principal components, which points
towards a random distribution of azimuths in the magnetic field.}
\label{principal_components}
\end{figure}

\section{Analysis and discussion}
We investigate how the distribution of polarization amplitudes changes with the spatial resolution
following two approaches. First, we compare the Hinode data set degraded to
different spatial resolutions by adding adjacent pixels. This allows us to study
the very same quiet Sun region at different spatial resolutions. In spite of the
lower spatial resolution, POLIS and ZIMPOL observations present a better signal-to-noise
ratio, thus allowing the detection of fainter signals. In order to complete the
picture, we compare all data sets, although it is fundamental to remind
that each observation represents the quiet Sun at different times and
in different regions of the solar surface.

The comparison of the polarization amplitudes of different data sets is not 
straightforward. As stated in \cite{marian_andres_08}, the presence of noise introduces
some systematic effects in the weak polarization tails of polarization 
amplitude histograms. In addition to purely instrumental errors, Stokes $V$ profiles 
are affected by asymmetries in the line profile, thus making it difficult to estimate
the amplitude. We propose a method based on principal component analysis (PCA) to
overcome all these issues and end up with sensible comparisons between
the polarization amplitudes of different data sets. Figure \ref{principal_components}
shows the first (solid line) and second (dashed line) principal components (PC)
of the Stokes $V$ (upper panel) and Stokes $Q$ and $U$ (lower panel).
In the case of Stokes $V$, the first PC (containing the vast majority of 
variance of the data set) is a typical antisymmetric Zeeman profile. For Stokes
$Q$ and $U$, the first PC is a typical symmetric Zeeman profile. The rest of the PC's
contain information about velocity shifts, broadenings, asymmetries, etc. \citep{skumanich_02}. 
Consequently, if we describe our data sets using only the
first PC, we fundamentally filter out all contributions except the one related
to the amplitude of the profiles. With this procedure, the contribution of 
noise and asymmetries is also strongly reduced. 

The amplitude of circular polarization ($A_V$) is computed as follows. 
We calculate the projection of the complete Stokes $V$ data set onto the first PC.
Then, since the first PC is antisymmetric, we define the amplitude of the Stokes $V$ profiles
as the semi-difference between the amplitudes of the blue and red lobes.
On the other hand, the linear polarization amplitude ($A_L$) is defined
as the maximum value of the quantity $(Q_1^2+U_1^2)^{1/2}$, $Q_1$ and $U_1$ being 
the projection of the data set in the first PC of Stokes $Q$ and $U$, respectively.
Interestingly, this simple exercise allows us to extract an important conclusion from the Stokes $Q$ and $U$
PC's. 
Figure~\ref{principal_components} shows that the first and second PC's for Stokes $Q$ and $U$
are nearly indistinguishable.
Moreover, higher PC's having a significant variance (not shown in the figure) behave similarly.
This indicates that the the azimuth of the magnetic field is uniformly distributed, with no
preferred direction.

\begin{figure*}[!t]
\includegraphics[width=0.5\textwidth]{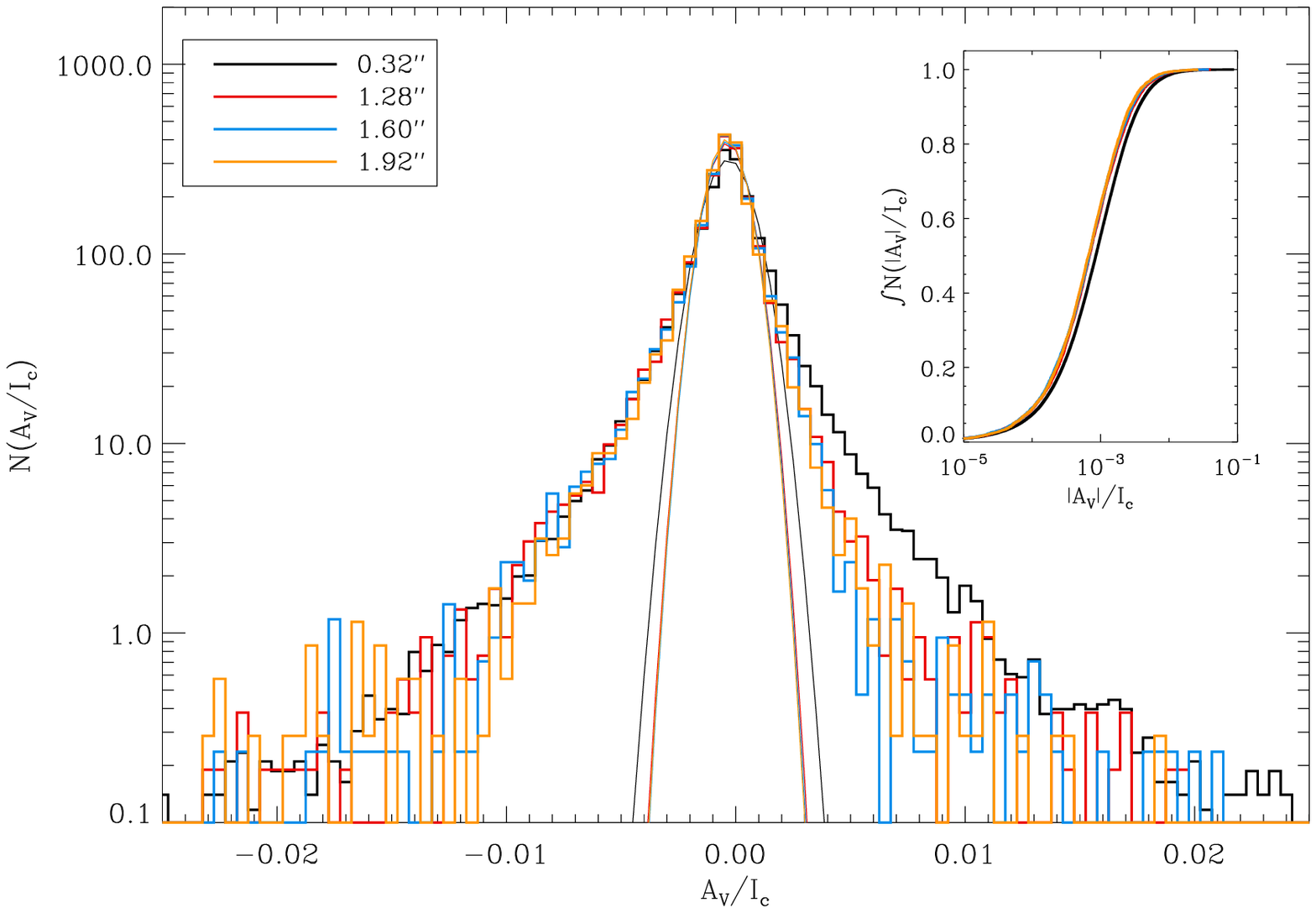}
\includegraphics[width=0.5\textwidth]{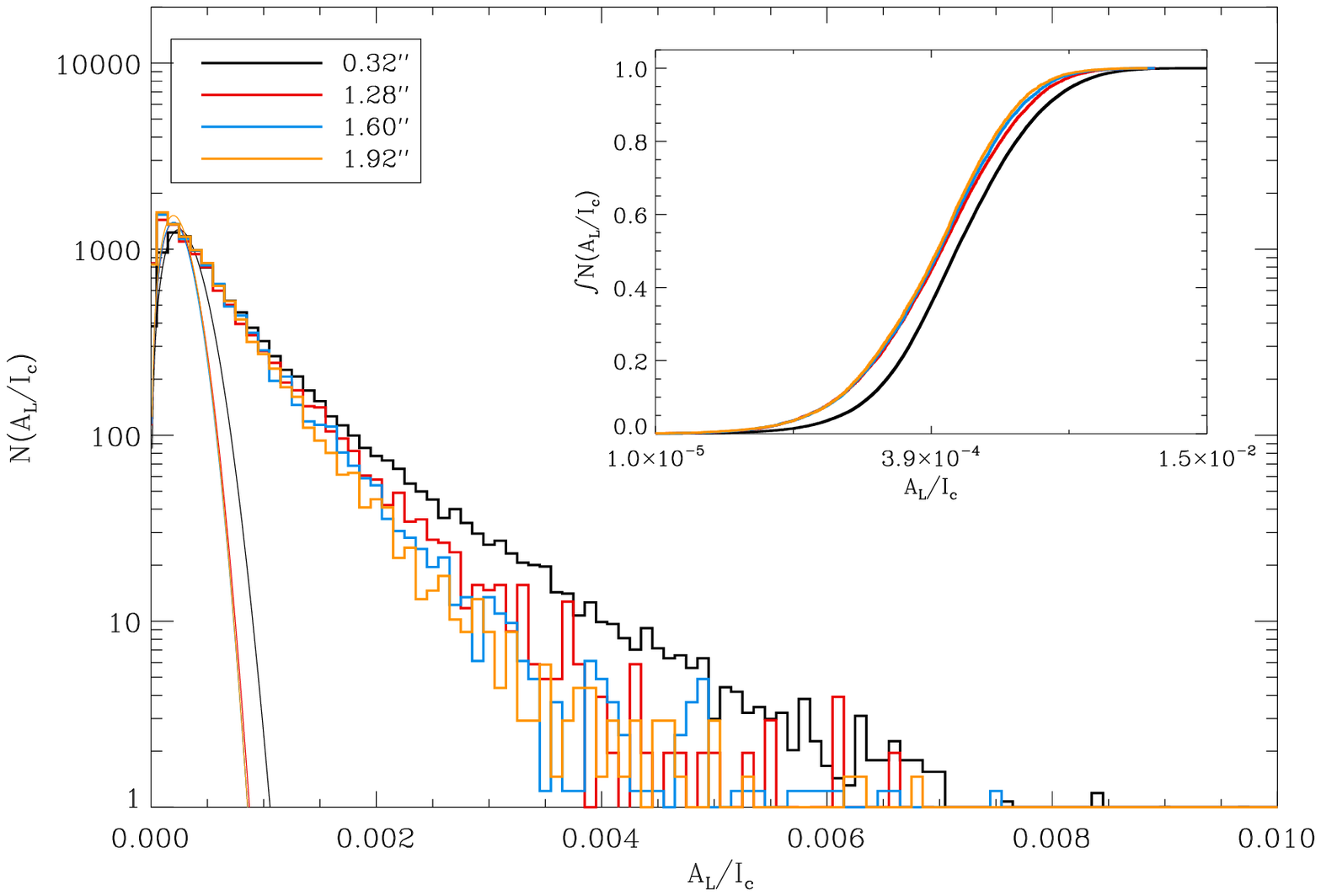}
\caption{Histograms of the circular (left panel) and linear polarization (right panel) of Hinode data
degraded to different spatial resolutions. The inset windows represent the empirical cumulative functions for all the histograms.
Gaussian fits to the central part of the $A_V$ histogram and the ensuing Rayleigh distributions for $A_L$
assuming isotropy are also shown.}
\label{hist_amplitudes_varia_res}
\end{figure*}

The left panel of Fig. \ref{hist_amplitudes_varia_res} shows the histograms of
$A_V$ for Hinode data from the highest possible spatial resolution of 0.32$''$ 
down to 1.92$''$, obtained by direct binning of the signals. Since adding up adjacent pixels reduces the noise of the data, we add 
a random noise so that all data sets at different spatial resolutions have the same noise level before computing the PC. The inset shows the
empirical cumulative distribution function for $A_V$. The
core of the circular polarization amplitude histogram changes marginally when
artificially degrading the data, while the tails are more extended for the highest 
resolution Hinode data. This can be understood if at $0.32''$ we are able to detect a
small amount of large polarization signals that is lost for lower resolutions. Note that the
histograms are plotted in log-linear scale, so that the difference detected in the
tails contribute negligibly to the total area of the histogram. 
In particular, as seen from the cumulative distribution functions, half of the area of the histograms is contained in
circular polarization amplitudes smaller than 8.4$\times 10^{-4} I_\mathrm{c}$ and 6.6$\times 10^{-4} I_\mathrm{c}$
for the 0.32$''$ and the lowest resolution data, respectively. This
indicates that the median (absolute value of $A_V$ at which we find
half of the total mass of the histogram) of the Stokes $V$ amplitudes has increased by just a factor of 1.27 when increasing the
spatial resolution from 1.92$''$ to 0.32$''$. 

Regarding the linear polarization histograms, the right panel of 
Fig.~\ref{hist_amplitudes_varia_res} shows that the mean values are 
5.2$\times 10^{-4} I_\mathrm{c}$ and 4.1$\times 10^{-4} I_\mathrm{c}$ for the 0.32$''$ and the low resolution data,
respectively. Accordingly, the median linear polarization signal increases by
a factor of 1.27 when increasing the spatial resolution from 1.92$''$ to 0.32$''$.

The previous values for the median of the circular and linear polarization
amplitudes indicate that the quiet Sun magnetism does not change much 
when increasing the spatial resolution from a few arcsec to sub-arcsec scales.
We now focus on the whole shape of the histograms. 
First, the circular polarization amplitude histograms do not present
a Gaussian shape throughout the full range of amplitudes, manifesting signatures of 
intermittency \citep[e.g.,][]{stenflo03,andres09}. Additionally, \cite{marian_andres_08} 
also found that these extended tails are modified when observing at different
heliocentric angles, an indication that these tails correspond to non-isotropic
structures. However, the core of the histograms 
can be well reproduced by a Gaussian function. The standard deviation found for the
different spatial resolutions considered are: $\sigma_{0.32}=1.0 \times 10^{-3} I_\mathrm{c}$, 
$\sigma_{1.28}=8.6\times 10^{-4} I_\mathrm{c}$, and $\sigma_{1.60}=\sigma_{1.92}=8.4 \times 10^{-4} I_\mathrm{c}$. 
This demonstrates that the core of the histograms at different spatial resolutions,
i. e., for amplitudes between $\pm 2.0 \times 10^{-3} I_\mathrm{c}$ 
are Gaussian and almost indistinguishable. The percentage of the observed area in the Sun covered 
by pixels with amplitudes in the core is 77\%, 82\%, and 84\% for the 0.32$''$, 1.28-1.60$''$ and 1.92$''$ data sets,
respectively. A deviation from the Gaussian shape occurs for pixels with larger circular
polarization amplitudes and they do display a trend with the spatial resolution.
Therefore, we can state that $\sim$20\% of the pixels show an increase in the
magnetic flux density when increasing the spatial resolution, a consequence of the
small number of signal cancellations at 0.32$''$.

It has been recently pointed out by \cite{marian_andres_08} and \cite{andres_09} that
the distribution of magnetic field vectors is close to isotropic in the 
internetwork. As a result, if the circular polarization signal and hence the longitudinal component of
the magnetic field ($B_z$) follows a Gaussian distribution, both $B_x$ and $B_y$
components necessarilly follow the same distribution. The amplitude of linear
polarization (as defined in this paper) under the weak field approximation
is proportional to $B_x^2+B_y^2$ and should follow an exponential distribution.
But this is true only if the fields are resolved, i.e., for individual Stokes $Q$ and
$U$ profiles. If we assume that the signal from each resolution element is
the addition of many Stokes $Q$ and $U$ profiles, their probability distribution should
converge to a Gaussian distribution accordingly the central limit theorem. Consequently,
the linear polarization signal should become close to a Rayleigh distribution. This
is compatible with the drop to zero of our observational histograms. 
For comparison, we overplot Rayleigh distributions with modes equal to
3.5-4 times the standard deviation of the Gaussian fits for $A_V$. This value
is in good accordance with the expected sensitivity difference for linear/circular
polarization for the 630 nm lines \citep{marian_08}.

It is evident that the smaller scale magnetic elements manifesting a Gaussian behavior
on the circular polarization histograms are identified with the lowest amplitudes of 
linear polarization. Moreover, the linear polarization histograms at different spatial 
resolutions behave similarly at the lowest signals, i. e., for amplitudes smaller 
than $10^{-3} I_\mathrm{c}$. The percentage of pixels below this threshold amounts
to 75\% for 0.32 $''$, 82\% for 1.28$''$, 84\% for 1.6$''$, and 86\% for 1.92$''$. 
The non-Rayleigh tails (20\% of the signals) are again more extended for the data sets
with higher resolution.

\begin{figure*}[!t]
\includegraphics[width=0.5\textwidth]{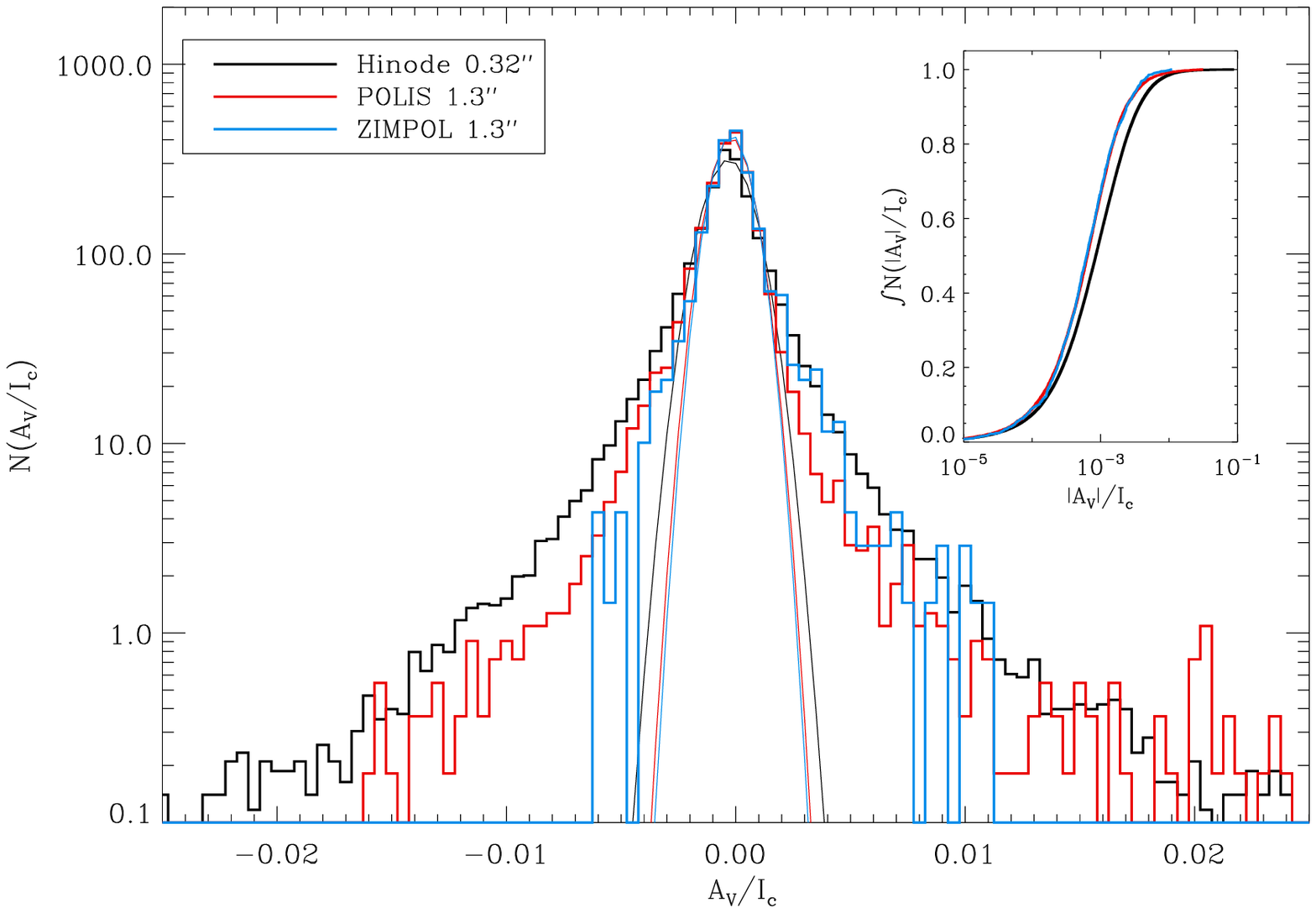}
\includegraphics[width=0.5\textwidth]{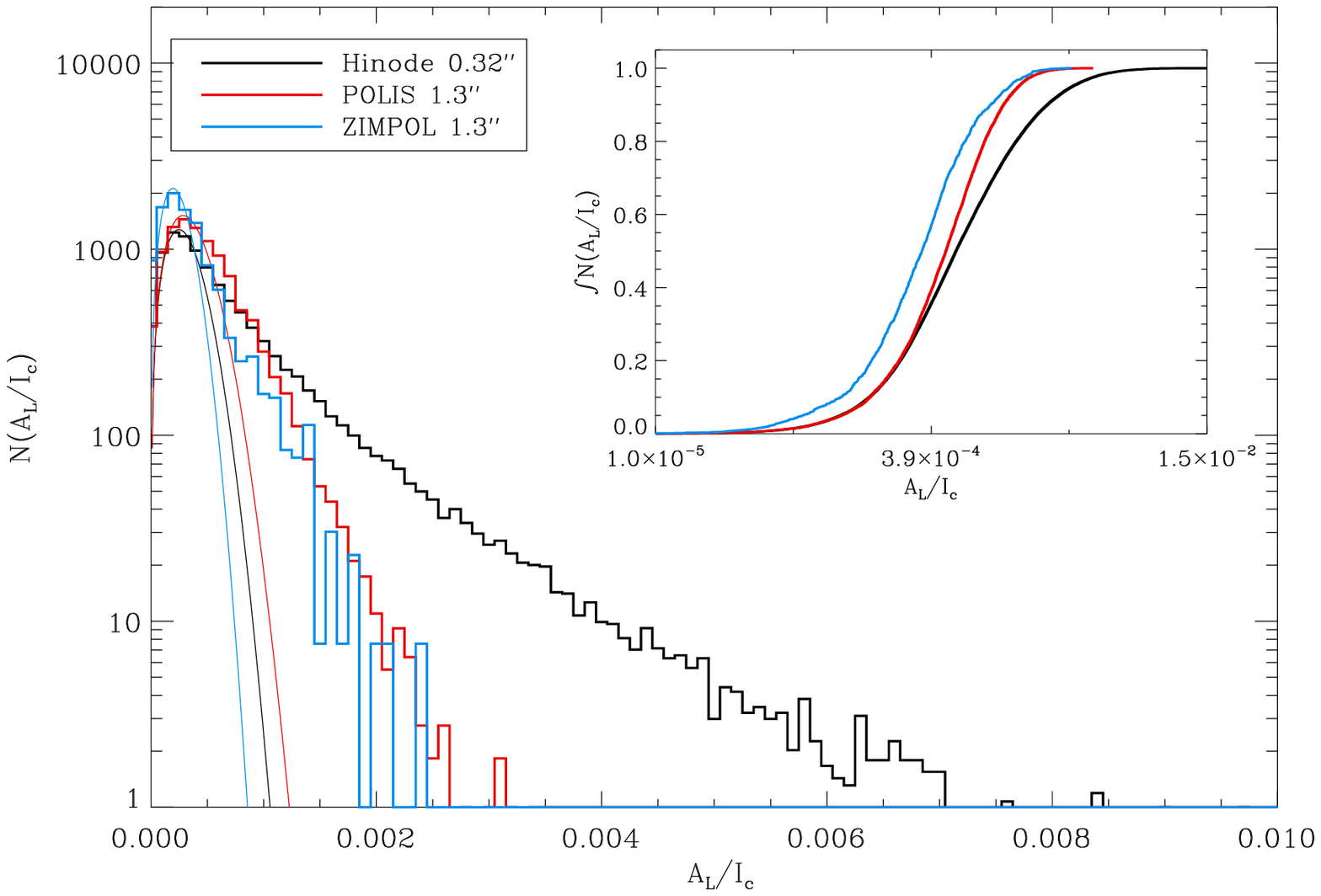}
\caption{Histograms of the circular (left panel) and linear polarization (right panel) of Hinode, POLIS,
and ZIMPOL data sets. The three observed data sets result in different spatial resolutions. The inset 
windows represent the empirical cumulative distribution functions for all the histograms.
Note that, due to the peculiarities of the ZIMPOL instrument, the
seeing-induced cross-talk effects are minimized with respect to the POLIS data and
gives more confident data, especially when mounted at TH\'EMIS.}
\label{hist_amplitudes}
\end{figure*}

We analyze in Fig. \ref{hist_amplitudes} the statistical properties of the polarization
signals for the different observational data sets. The median 
of $|A_V|$ for Hinode is a factor 1.35 larger than for ZIMPOL, while $A_L$ is
a factor 1.67 larger. Analogously, the mean unsigned flux changes a factor 2, from 2.2$\times 10^{-3} I_\mathrm{c}$ to 1.0-1.1$\times 10^{-3} I_\mathrm{c}$, when the spatial resolution is degraded from the HINODE one to the POLIS and ZIMPOL ones. This is in agreement with the values published in the literature \citep[see compilation in Fig. 1 of][]{jorge_09}.
Taking into account that the ZIMPOL data contains less observed points 
and that the observations are taken at different days, we reckon that the increase on
polarization amplitudes is in agreement with the analysis of Hinode at different spatial resolutions.
Moreover, we also behold that most of the signals are represented by a Gaussian
(Rayleigh) core for circular (linear) polarization that is insensitive to
spatial resolution.

These observed trends with the spatial resolution are strong constraints to the
nature of the magnetic fields in the internetwork. Any proposed model for the
quiet Sun magnetism should reproduce the change on the Stokes $Q$, $U$ and $V$
amplitudes with spatial resolution. For instance, the single magnetic
atmosphere surrounded by a field-free component and the simple microturbulent model
are not valid models for the internetwork magnetism. The former because
a single magnetic element observed at increasingly better resolutions should induce
a dramatic increase in the polarimetric signal; while the latter because a microturbulent
scenario is not compatible with the extended tails that evidence a cascade of
spatial scales coexisting in the internetwork quiet Sun. The correct interpretation of the
quiet Sun should take into account that the magnetic field vector $\mathbf{B}$ depends on 
the scale $r$ and the description has to rely on the complete scale-dependent continuous probability 
distribution function $p(\mathbf{B}_1,r_1;\mathbf{B}_2,r_2;\ldots;\mathbf{B}_n,r_n)$ 
\citep[e.g.,][]{andres09}. If the scale of organization is much smaller than the resolution element, 
a large degree of cancellations occurs and the distribution of observed polarization
amplitudes quickly tends to a Gaussian distribution. Therefore, the Gaussian core of the
circular polarization amplitude histogram can be identified with the fields organized
well below the resolution element. The extended tails are thus identified with magnetic elements
organized at a larger scale. 
We calculate a ``characteristic size'' of the smallest magnetic elements through inversion with a model atmosphere that is formed, in each resolution element, by an isotropic distribution of magnetic field vectors all of them of the same size. With this model we get an upper limit to the size of these magnetic elements $\sim$10 km, which is smaller than the photon mean free path in the photosphere. Such small scale structuring of the quiet Sun magnetism was conjectured by \cite{jorge_egidio_valentin_96} based on the ubiquity of the Stokes asymmetries.However, although the majority of magnetic fields in the quiet Sun is organized at such small scales, models taking into account the cascade of scales are necessary to understand the quiet Sun magnetism.


The direct analysis of observables reveals that there is not a substantial
change in the polarization amplitudes from 1.3$''$ to $0.32''$. 
A fraction of 80\% of the circular (linear) signals follow a Gaussian (Rayleigh) 
distribution and seem to be insensitive to the spatial resolution. This important 
fact reveals that the Zeeman effect is indeed sensitive to the microturbulent magnetic field 
usually diagnosed with the Hanle effect. In fact, even a thousand of mixed-polarity, magnetic elements in the 
resolution element gives a detectable Zeeman signal \citep{arturo07}. The remainder 20\% of the observed polarization 
signals show a modification with spatial resolution, evidencing a cascade of spatial scales in the
internetwork. The very quiet Sun magnetism is thus very complex and we are forced to employ
sophisticated statistical modeling tools to infer its properties.

\begin{acknowledgements}
We acknowledge financial support from the Spanish MICINN through the project AYA2007-63881. Hinode is a Japanese mission developed and launched by ISAS/JAXA, with NAOJ as a domestic partner, and NASA and STFC (UK)
as international partners. It is operated by these agencies in cooperation with ESA and NSC (Norway).
\end{acknowledgements}


\end{document}